\documentclass[aps,prl,preprint,amsmath,amssymb,groupedaddress]{revtex4}

\usepackage{graphicx}
\usepackage{dcolumn}
\usepackage{bm}

\begin{document}

\title{Microwave-Induced Oscillations in the Magnetocapacitance:
Direct Evidence for Non-equilibrium Occupation of Electronic States}

\author{S.\,I.~Dorozhkin}
\author{A.\,A.~Kapustin}
\affiliation{Institute of Solid State Physics RAS, 142432
Chernogolovka, Moscow district, Russia}
\author{V.~Umansky}
\affiliation{Department of Physics, Weizmann Institute of Science, 76100 Rehovot, Israel}
\author{K.~von~Klitzing and J.\,H.~Smet}
\affiliation{Max-Planck-Institute f\"ur Festk\"orperforschung, Heisenbergstrasse 1, D-70569 Stuttgart, Germany}

\date{\today}

\begin{abstract}
In a two-dimensional electron system, microwave radiation may
induce giant resistance oscillations. Their origin has been
debated controversially and numerous mechanisms based on very
different physical phenomena have been invoked. However none of
them have been unambiguously experimentally identified, since they
produce similar effects in transport studies. The capacitance of a
two-subband system is sensitive to a redistribution of electrons
over energy states, since it entails a shift of the electron
charge perpendicular to the plane. In such a system microwave
induced magnetocapacitance oscillations have been observed. They
can only be accounted for by an electron distribution function
oscillating with energy due to Landau quantization, one of the
quantum mechanisms proposed for the resistance oscillations.
\end{abstract}

\maketitle

Recent studies of non-equilibrium phenomena in two-dimensional
electron systems (2DES) exposed to microwave (MW) radiation have
revealed remarkable transport effects. The most prominent one is
the appearance of giant microwave-induced magnetoresistance
oscillations (MIRO)~\cite{Zudov1,Ye}. In the main minima the
resistance may approach zero and so-called zero resistance states
develop~\cite{Mani,Zudov2,Dorozh2015}. MIRO have proved to be a
general effect as they by now have been observed in both
degenerate 2DES, such as in GaAs/AlGaAs~\cite{Zudov1,Ye},
Si/SiGe~\cite{ZudovSi} and ZnO/MgZnO~\cite{ZnO} heterostructures,
as well as in non-degenerate 2D electron gases on liquid
helium~\cite{KonstHe} (for additional references we refer to the
review~\cite{DmitrievRMP}). MIRO are periodic in
$\omega/\omega_{\rm c}$, where $\omega/2\pi$ is the MW frequency
and $\omega_{\rm c}=eB/m^*$ is the cyclotron frequency. The
proposed explanations of MIRO are based on classical or quantum
effects as reviewed in Ref.~\cite{DmitrievRMP} (see also a recent
classical theory~\cite{DyakonovPRL16}). Despite significant
progress and a satisfactory description of much of the
phenomenology by some of the proposed mechanisms, the theoretical
dispute has not been settled since crucial unresolved issues
remain. In particular, the absence of a dependence of MIRO on the
MW circular polarization direction~\cite{Smet,Ganichev} continues
to stimulate new theoretical ideas involving edge~\cite{Shep} and
contact~\cite{Mikhailov} phenomena and invigorates the
debate~\cite{ZudovComm,MikhailovComm}. Magnetotransport
experiments alone are likely insufficient to identify
unambiguously all mechanisms  active in experiment. Apart from
MIRO, oscillations of the same periodicity were also found
elsewhere in other electronic transport properties. Under MW
radiation voltages and currents develop even in the absence of
external sources. These photo-galvanic
signals~\cite{Willett,Bykov,Dorozh09} can be traced back to MIRO
and the MW induced changes in the
dc-conductivity~\cite{PRB09,PRB11}. Therefore they too are of
limited use to reduce the clutter. There is clearly a strong need
to look at other physical quantities than the conductivity.

Here, using the magnetocapacitance technique and a suitable sample
design we demonstrate that MW radiation generates a
non-equilibrium distribution of electrons among the Landau levels
oscillating with energy. Such a non-equilibrium distribution
function~\cite{Dorozh03,Dmitriev03,Dmitriev05,Raichev} represents
one of the most elaborate theoretical pictures to account for MIRO
as a bulk quantum phenomenon. While capacitive measurements in the
presence of the MW radiation have been reported for quasi-2D
electrons above a He surface ~\cite{Abdur,Chep}, in degenerate
2DES the capacitance has not been addressed yet in this microwave
context. It has also been widely used for studying the equilibrium
2DES compressibility~\cite{Stiles}. The sample consists of a field
effect transistor with a back gate and an electron channel that
resides in an asymmetric, wide GaAs quantum well (QW) with two
occupied subbands. A microwave induced redistribution of the
electrons along the energy scale modifies the occupation of both
subbands and is accompanied by a shift of the electrons {\em
perpendicular to the QW plane}. This shift alters the capacitance
which is primarily sensitive to the occupation of the highest
subband whose wave function is located closer to the gate. Hence,
the capacitance can capture directly microwave induced
oscillations in the electron energy distribution. The sensitivity
to vertical electron shifts is what distinguishes this capacitance
measurement from all previous transport experiments. A second
frequently invoked
mechanism~\cite{Ryzhii1,Ryzhii2,Durst,Ryzhii3,Aleiner,Dmitriev09}
to explain MIRO involves a microwave induced impurity scattering
assisted displacement of the electrons {\em in the plane}. The
equations describing MIRO within this displacement model are,
except for a temperature dependent prefactor, identical to the
equations obtained from a picture based on a non-equilibrium
energy distribution function~\cite{Dmitriev05,Dmitriev09}. An
unambiguous separation of these two possible contributions is
therefore non-trivial. The capacitance measurements can not
exclude that the displacement mechanism is active in MIRO as they
are not sensitive to lateral displacements. However, they can
without any ambiguity prove the formation of a non-trivial
electron energy distribution function by the microwaves and
thereby provide support for a bulk quantum origin of MIRO.

We measured two identical Hall bar samples processed side by side
on the same piece of a GaAs/AlGaAs heterostructure. The electron
system resides in a 60 nm wide GaAs QW. An in-situ grown back gate
allows to tune density and measure the capacitance. In these
samples the second subband gets populated at total electron
density $n_{\rm s}\approx 1.8\times 10^{11}\ {\rm cm}^{-2}$ for a
gate voltage $V_{\rm g} > 0.15\,{\rm V}$, as will be shown below.
Further experimental details are deferred to the Supplemental
Material~\cite{Supp}. The samples were placed in a stainless steel
tube with a diameter of 18 mm. It served as an oversized waveguide
for the MW radiation whose frequency was varied from 54 to 78 GHz.
The measurements were performed in a pumped liquid $^3{\rm He}$ at
0.5 K.

Fig.~1 illustrates the behavior of the longitudinal resistivity,
$\rho_{\rm xx}$ (panels b and c), and variation of the capacitance
$\Delta C$ (panels a and d),in the absence and presence of MW
radiation for two gate voltages: $V_{\rm g}=0$ and 1~V. As shown
below, these  voltages correspond to one and two occupied
subbands, respectively. While the $\rho_{\rm xx}$ curves (panels b
and c) look very similar in both regimes, the magnetocapacitance
traces in panels a and d are qualitatively different. A close
inspection allows to identify four types of $1/B$-periodic
oscillations. At high $B$-fields all $\rho_{\rm xx}$ traces
exhibit the well-known Shubnikov-de Haas (SdH) oscillations. The
microwaves induce additional, large oscillations in the
magnetoresistivity. These are just MIRO. The oscillations in the
magnetocapacitance measured in the absence of radiation reflect
the oscillations in the thermodynamic density of states (DOS),
$\partial n_{\rm s}/\partial\mu$~\cite{Stiles}, brought about by
Landau quantization. Here $\mu$ is the chemical potential of the
2DES. The MW radiation suppresses the amplitude of these DOS
oscillations in the magnetocapacitance. In the density regime with
only one occupied subband (regime I) shown in panel d, the
suppression is the only effect of the radiation on the
capacitance. In contrast, when two subbands are occupied (regime
II) the MW radiation induces additional oscillations with a new
period. These microwave induced capacitance oscillations,
hereafter referred to as MICO, have been demarcated in panel a.
They exhibit a node at the same $B$-field where the MIRO have
their rightmost zero, i.e., when $\omega = \omega_{\rm c}$. This
is highlighted by the dashed line in Fig.1. It corresponds to the
cyclotron resonance (CR) of electrons with effective mass
$m^*=0.061\ m_{\rm e}$. This value of $m^*$  is close to the
$0.059\ m_{\rm e}$ recently obtained~\cite{Zudovm} from the MIRO
periodicity. The observation of MICO is the key experimental
result of this Letter.

In Fig.2a MICO have been plotted as a function of $1/B$ for three
different MW frequencies to highlight the following: (i) The
oscillations are indeed periodic in $1/B$ and the period $1/B_0$
does not depend on the frequency. (ii) The oscillation amplitude
indeed reveals a beating pattern with the leftmost node located at
the CR (for $m^*=0.061\ m_{\rm e}$). (iii) When crossing a node
the phase of the oscillations jumps by $\pi$. This can be seen
most easily with the help of the top axis. The abscissa is
obtained by normalizing $1/B$ with the oscillation period: $B_{\rm
0}/B$. Integer values initially correspond to maxima. However,
when the CR node is crossed, integer values align with minima
instead. In the topmost curve for 78 GHz radiation, a second node
at $\omega/\omega_{\rm c}=3/2$ is observed. When it is crossed,
integer values correspond again to maxima. The data in panel b
have been recorded at the same frequency as the bottom trace in
panel a, but for a different $V_{\rm g}$. A comparison unveils
that the MICO periodicity has changed while the node stays at the
CR position.

To identify the origin of MICO, it is instrumental to
systematically vary $V_{\rm g}$, monitor changes in the
oscillation period and compare the results for different
oscillation types. The outcome of such a study is summarized in
Fig. 3a. The SdH and DOS oscillations, also observed under
equilibrium conditions, help to extract the density and identify
when a second subband gets populated. At $V_{\rm g} = 0$ where
only one subband is occupied, the SdH and DOS oscillations have
the same period and phase as seen in panels c and d of Fig. 1
(short vertical lines). Minima appear when the Fermi level is
located within a cyclotron gap and an integer number $n$ of spin
degenerate Landau levels is occupied. This occurs at $B$-fields
for which $n_{\rm s1}=2nN_0$ with $n = 1,2,\ldots$, $N_0=eB/h$ the
Landau level degeneracy per spin and $n_{s1}$ the density in the
lowest subband. It results in a 1/B-periodicity equal to
$2e/hn_{s1}$ from which $n_{s1}$ can be calculated. However, more
generally, this expression can be used to convert any observed
periodicity into a density whose meaning needs to be interpreted
properly. Hereafter, the densities extracted from SdH and  DOS
oscillations as well as MICO will be denoted as $n_{l}$ with
subscript $l =$ SdH, DOS or MICO. They have been plotted in Fig.
3a together with $n_{\rm H}$ deduced from the Hall resistance at
low $B$. The latter increases with $V_{\rm g}$ at a constant rate.
When tracing $n_{\rm SdH}$ and $n_{\rm DOS}$ to positive $V_{\rm
g}$ in Fig. 3a, we note that their behavior is very different. For
instance, $n_{\rm SdH}$ remains approximately constant. Referring
to the raw data recorded at $V_{\rm g} = 1.0\ {\rm V}$ (Fig. 1b),
the envelope of the SdH oscillation pattern has become more
complicated but its main period indeed remains close to that at
$V_{\rm g}=0$. On the other hand, $n_{\rm DOS}$ drops considerably
when $V_{\rm g}$ exceeds 0.15 V. This is also apparent in the raw
data of Fig. 1 where the horizontal arrows in panels a and b mark
the oscillation periods. In Fig. 3a also the sum  $n_{\rm SdH} +
n_{\rm DOS}$ has been plotted. It coincides with $n_{\rm H}$ for
$V_{\rm g} > 0.15\ V$. In regime I, all densities are equal:
$n_{\rm SdH}=n_{\rm DOS}=n_{\rm H}=n_{\rm s} = n_{\rm s1}$. We
assert that all these observations can be understood
straightforwardly assuming the second subband becomes occupied for
$V_g > 0.15\ {\rm V}$.

In a wide QW with an asymmetric potential profile, subband wave
functions are located at different distances from the gate
effectively mimicking a bilayer system as schematically
illustrated in panel c of Fig.~3. For such a case, a variation of
$V_{\rm g}$ primarily changes the density in the second subband or
layer 2 closest to the gate and only slightly affects the charge
in remote layer 1 (the first subband). Then, the DOS oscillations
are determined by Landau quantization in the second subband and
their periodicity is governed by the density in this subband,
$n_{\rm s2}$, only (for more details see
Refs.~\cite{Dolgopolov,Dorozh2016}). Back to the SdH oscillations,
one may expect two sets of oscillations, determined by carrier
densities in both subbands, $n_{\rm s1}$ and $n_{\rm s2}$. In our
raw data the fastest oscillations are however easiest to discern.
They are associated with Landau quantization of the lowest subband
with the largest population, $n_{\rm s1}$, which remains
approximately fixed since the gate electric field is screened by
the electrons in the second subband. These electrons generate
weaker oscillations in the envelope of the rapid oscillations from
the first subband. We note that the linear $n_{\rm s2}(V_{\rm g})$
dependence as well as nearly constant value of $n_{\rm s1}$ shown
in Fig.3(a) are similar to those reported in other studies of
unbalanced bilayer electron systems (see, for example,
Refs.~\cite{Stoermer,Shayegan} for single and double quantum
wells, respectively). The Hall density $n_{\rm H}$ corresponds to
the total density $n_{\rm s}$. This interpretation of the data is
strongly supported by the experimentally established relation
$n_{\rm H}=n_{\rm DOS}+n_{\rm SdH}$. Then we finally conclude that
$n_{\rm DOS}=n_{\rm s2}$ and $n_{\rm SdH}=n_{\rm s1}$. Fig. 3a
also contains the density extracted from MICO, $n_{\rm MICO}$.
Apparently it is identical to $n_{\rm SdH}-n_{\rm DOS}$, which in
view of the above discussion is equivalent to the subband
population difference: $n_{\rm MICO}= n_{\rm s1}-n_{\rm s2}$.

Clearly, the response to radiation is very different in $\rho_{\rm
xx}$ and the magnetocapacitance. It follows that MICO cannot be
explained in terms of a MW induced variation of the conductivity.
The extracted density from MICO points to its origin, since the
condition $n_{\rm s1}-n_{\rm s2}=2nN_0$ is equivalent to
$\Delta\equiv \varepsilon_2-\varepsilon_1=n\hbar\omega_{\rm c}$
where $\varepsilon_{\rm j}$ ($j=1,2$) are the subband energies. At
this commensurability condition, Landau levels of the two subbands
are aligned. We therefore argue that MICO reflect a MW induced
charge redistribution among the two subbands whose magnitude
oscillates with $B$. This is detected in the capacitance since, in
our sample, it selectively responds to occupation of the second
subband. This interpretation is further corroborated by the
capacitance step observed in Fig. 3a at $V_{\rm g}\approx
0.15$\,V. This step is caused by occupation of the second subband
(i.e., formation of the second layer) with a center of mass of the
wave function located approximately 20 nm closer to the gate than
that of the 2DES in regime I at $V_{\rm g}\lesssim 0.15$\,V
(compare Figs 3b and 3c). It demonstrates sensitivity of our
measurements to variation of the charge distribution in the QW.
For the sake of completeness, we note that also
magnetointersubband oscillations (MISO) with a period determined
by the relation $\Delta=n\,\hbar\omega_{\rm c}$ may occur in the
magnetoresistance due to intersubband
scattering~\cite{Polyan,SST1,SST2,PRB46,Raikh}. They can be
strongly affected by radiation, which may introduce
nodes~\cite{Bykov2,Raichev}. They can be explained by both
non-equilibrium distribution function and displacement
mechanisms~\cite{Raichev,Raichev2}. At 0.5 K the MISO are masked
by the SdH oscillations.

To substantiate our assertion that the magnetocapacitance
oscillations prove that microwaves create a non-equilibrium
distribution function oscillating with energy due to Landau
quantization, we have analyzed our results within a distribution
function model generalized to the case of two occupied
subbands~\cite{Raichev}. The equation for the MW induced
correction $\delta f(\varepsilon)$ to the Fermi distribution
function $f_{\rm F} (\varepsilon)$ in a balanced double-quantum
well structure reads as follows~\cite{Raichev}:
\begin{equation}
\delta f(\varepsilon) \simeq \frac{\hbar\omega}{2} \frac{\partial
f_{\rm F}}{\partial\varepsilon} P_{\rm
\omega}\sin\frac{2\pi\omega}{\omega_{\rm c}}\sum_{j=1,2} d_{\rm j}
\sin\frac{2\pi(\varepsilon-\varepsilon_{\rm j})}{\hbar\omega_{\rm
c}}.
\end{equation}
\noindent The  dimensionless factor $P_{\rm \omega}$ is
proportional to the MW power absorbed by the 2DES. This equation is
derived to first order with respect to the small Dingle
factors $d_{\rm j}=\exp(-\pi/\omega_{\rm c}\tau_{\rm j})$ for each
subband and under the assumptions that $\hbar \omega\ll kT \ll \varepsilon_{\rm
F}-\varepsilon_{\rm j}$. Here, $\tau_{\rm j}$ is the electron
quantum lifetime in the $j$-th subband and $\varepsilon_{\rm F}$
is the Fermi energy. Within this framework of approximations, the density of states
in a subband is given by
\begin{equation}
D_{\rm j}(\varepsilon)=\frac{m^*}{\pi \hbar^2} \left[1-d_{\rm
j}\cos\frac{2\pi(\varepsilon-\varepsilon_{\rm
j})}{\hbar\omega_{\rm c}}\right].
\end{equation}
The MW induced variation of the density in the first subband
($\varepsilon_1<\varepsilon_2$) is equal to
\begin{equation}
\delta n_{\rm s1}=-\delta n_{\rm s2}=\int_{\varepsilon_1}^{\infty}
D_1(\varepsilon)\delta
f(\varepsilon)\,d\varepsilon=-d_1d_2\frac{m^*}{\pi\hbar^2}\frac{\hbar\omega}{4}
P_{\rm \omega}\sin\frac{2\pi\omega}{\omega_{\rm
c}}\sin\frac{2\pi(\varepsilon_2-\varepsilon_1)}{\hbar \omega_{\rm
c}}.
\end{equation}
\noindent When $\varepsilon_2-\varepsilon_1=\Delta\gg
\hbar\omega$, Eq.(3) describes magnetooscillations with a
periodicity determined by the commensurability between the
cyclotron energy and the subband spacing: $\Delta=n\,\hbar
\omega_{\rm c}$. The beating pattern and nodes are caused by the
factor $\sin(2\pi\omega/\omega_{\rm c})$. The nodes are located at
$\omega/\omega_{\rm c}=(n+1)/2$. This oscillation pattern
described by Eq.(3) matches all the observed features of MICO.
These oscillations should also persist in unbalanced system in
which the center of mass of the wave functions of the two subbands
are spatially separated. Then the oscillating redistribution of
electrons between the subbands (i.e., the layers) $\delta n_{\rm
s1}=-\delta n_{\rm s2}$, produces oscillations in the capacitance.
This accounts for our experimental observations.

In summary, by implementing a new experimental approach to study
non-equilibrium phenomena in 2DES we have discovered microwave
induced magnetooscillations of an electrical capacitance. We have
shown that these oscillations reflect redistribution of electrons
between two occupied subbands which oscillates with magnetic field
due to non-trivial distribution of electrons among Landau levels.
Our observation establishes unequivocally the importance of this
non-equilibrium distribution function scenario which was developed
to explain MIRO.

We acknowledge fruitful discussions with I. A. Dmitriev.
Experiment and data evaluation of this work were supported by the
Russian Scientific Foundation (Grant 14-12-00599). JHS and VU
acknowledge support from the GIF.

\begin{figure}[tbh!]
\centering {\includegraphics{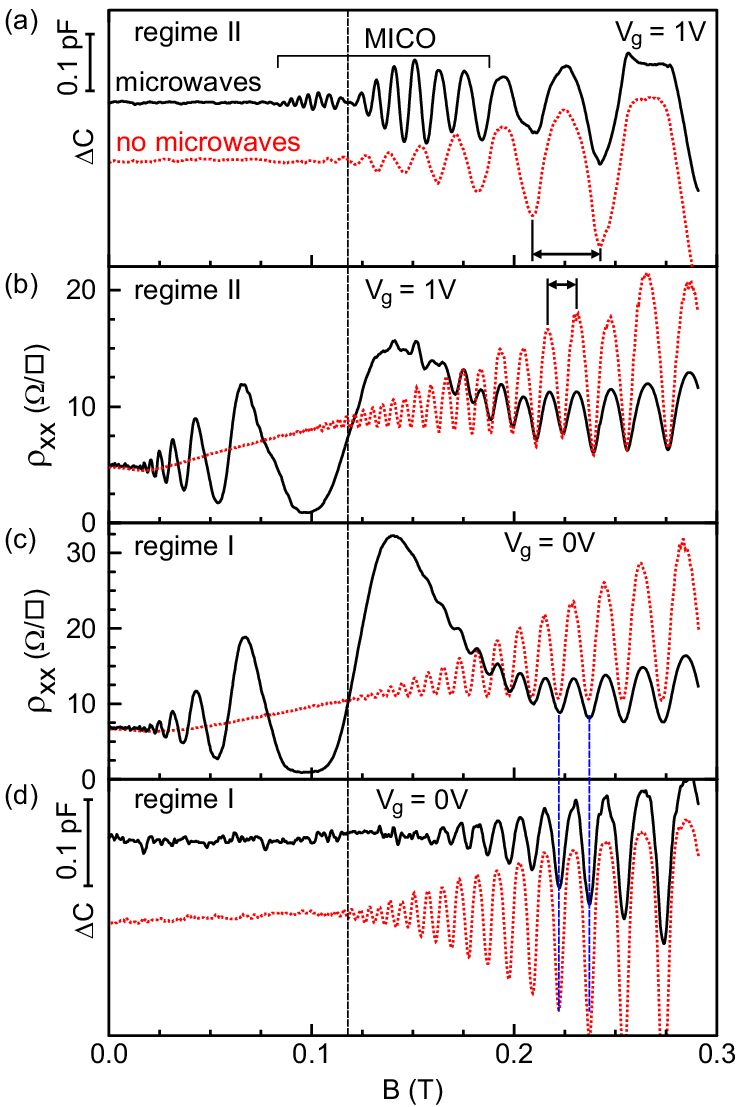}} \caption{
Magnetocapacitance variation $\Delta C$ (panels (a) and (d)) and
magnetoresistivity $\rho_{\rm xx}$ (panels (b) and (c)) under MW
radiation (solid lines) and without radiation (dashed lines). For
the sake of clarity the dashed magnetocapacitance curves are shifted
down by 0.1 pF relative to the solid ones. The data for the two
occupied subbands (regime II) are shown in panels (a) and (b) and
for one occupied subband (regime I) in panels (c) and (d). MW
frequency $\omega/2\pi=54$~GHz. The short vertical lines in panels
(c) and (d) are drawn through the oscillation minima in $\rho_{\rm
xx}$ and $\Delta C$.}
\end{figure}

\begin{figure}[tbh!]
\centering {\includegraphics{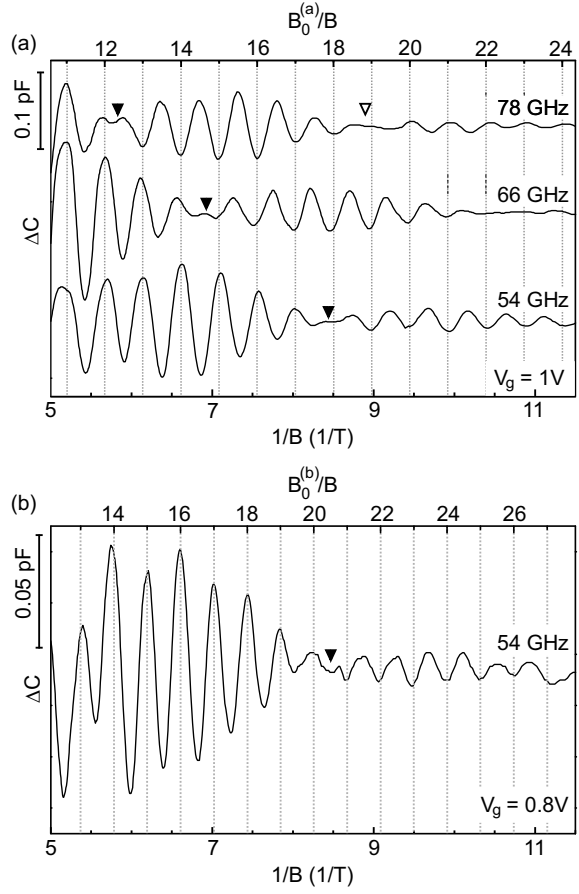}} \caption{MICO as a
function of inverse magnetic field for different MW frequencies
(panel(a)) and gate voltages (panels (a) and (b)) shown at the
curves. The solid (open) arrows mark the points where
$\omega/\omega_{\rm c}=1$ (3/2) calculated for $m^*=0.061m_{\rm e}$.
The integer values on the upper scales correspond to the oscillation
numbers. Here $1/B_0^{\rm (a)}$ and $1/B_0^{\rm (b)}$ are the
oscillation periods for data shown in panels (a) and (b),
respectively.}
\end{figure}

\begin{figure}[tbh!]
\centering {\includegraphics{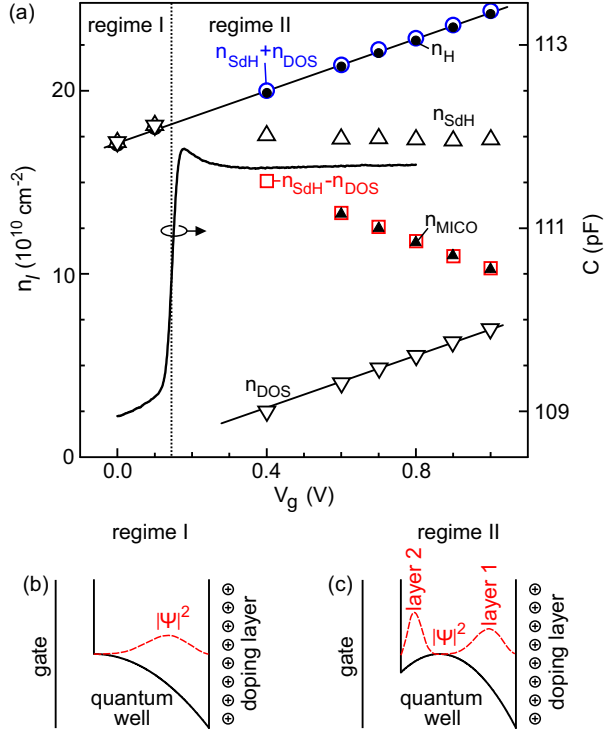}} \caption{(a) Gate
voltage dependencies of electron densities $n_l$ determined from 0.5
K SdH oscillations ($n_{\rm SdH}$, upward open triangles),
magnetooscillations of dark capacitance ($n_{\rm DOS}$, downward
open triangles), MICO ($n_{\rm MICO}$, upward close triangles), and
Hall resistance ($n_{\rm H}$, solid dots). The vertical dotted line
separates regions I and II with one and two occupied subbands,
respectively. The calculated values of $n_{\rm DOS}+n_{\rm SdH}$
(open circles) and $n_{\rm SdH}-n_{\rm DOS}$ (open squares) are
shown for region II and $V_{\rm g}\geq 0.4$~V. The lower thin line
is drawn through the $n_{\rm DOS}$ data points parallel to that
corresponding to the $n_{\rm H}(V_{\rm g})$ dependence. The thick
solid curve is the experimental dependence of capacitance versus
gate voltage at $B=0$. Potential and electron density distribution
$|\Psi|^2$ in the QW are schematically shown in panels (b) and (c)
for one and two occupied subbands, respectively.}
\end{figure}

\clearpage

\noindent {\bf Supplemental Material for "Microwave-Induced
Oscillations in the Magnetocapacitance: Direct Evidence for
Non-equilibrium Occupation of Electronic States"}
\\
\\
S.~I.~Dorozhkin$^1$, A.~A.~Kapustin$^1$, V. Umansky$^2$,
K.~von~Klitzing$^3$ and J.~H.~Smet$^3$
\\
\\
\textit{$^1$Institute of Solid State Physics, Chernogolovka,
Moscow district, 142432, Russia\\
$^2$Department of Physics, Weizmann Institute of
Science, 76100 Rehovot, Israel\\
$^3$Max-Planck-Institut f\"{u}r Festk\"{o}rperforschung,
Heisenbergstrasse 1, D-70569 Stuttgart, Germany}
\\
\\
\noindent {\bf Sample design and experimental technique}\\
\ \\
\noindent Our magnetocapacitance and magnetoresistance studies have
been performed on two identical Hall bar samples processed side by
side on the same piece of a GaAs/AlGaAs heterostructure. In this
structure, the electron system resides in a 60 nm wide GaAs quantum
well (QW). An in-situ grown homogeneously doped GaAs layer located
at a distance of 850 nm below the QW serves as the gate. The QW is
filled by electrons via modulation doping from a doped region in the
top AlGaAs layer at a distance of 65 nm from the QW. After cooling
down to 4.2 K, the samples were illuminated with a red LED till
saturation of the electron density. By changing the gate voltage
$V_{\rm g}$ the electron density $n_{\rm s}$ was varied in the
$(1.7-2.4)\times 10^{11}\ {\rm cm}^{-2}$ range. The electron
mobility remained above $4\times 10^6\ {\rm cm}^2/{\rm V\,s}$. In
these samples only one subband is occupied at $V_{\rm g} < 0.15$\,V
($n_{\rm s}<1.8\times 10^{11}\ {\rm cm}^{-2}$). The second subband
gets populated above these values. Electrons of this subband are
located closer to the gate than electrons of the first subband. The
Hall bar geometry consists of a 1.8 mm long and 0.4 mm wide mesa.
The distance between neighboring potential probes is 0.4 mm and 0.8
mm. The magnetotransport data were recorded with a standard lock-in
technique using an ac excitation of 200 nA at a frequency of 13 Hz.
To measure the capacitance between the gate and the 2DES, an 118 Hz
ac voltage with an rms amplitude of $20\,{\rm mV}$ was applied to
the gate. To minimize the stray capacitance, the gate was connected
to a coaxial cable. An SR570 current preamplifier was connected to
the source contact of the sample. The ac output voltage from the
preamplifier was amplified further with the input stage of an SR830
lock-in amplifier and both the in phase (active), as well as the
$90^{\rm o}$ out of phase (reactive) components of the signal were
measured. In this measurement configuration the latter component is
dominant and proportional to the capacitance. Both samples showed
identical dependencies of the capacitance on the magnetic field and
the gate voltage. The samples were placed in an oversized waveguide,
stainless tube of 18 mm diameter, and immersed in liquid $^3{\rm
He}$. The measurements were carried out at a base temperature of
approximately 0.5 K obtained by pumping $^3{\rm He}$ vapor. To study
MW photoresponse the samples were irradiated by radiation of the
54-78 GHz. Estimated MW power in the waveguide at the sample
location was 0.2 mW.

\end{document}